\documentclass[runningheads]{llncs}
\usepackage[T1]{fontenc}
\usepackage{graphicx,verbatim}
\usepackage{xcolor}

\usepackage{amssymb}
\usepackage{amsmath}
\usepackage{booktabs}

\makeatletter
\newcommand{\printfnsymbol}[1]{\textsuperscript{\@fnsymbol{#1}}}
\makeatother

\begin{document}
\title{Investigating a Policy-Based Formulation for Endoscopic Camera Pose Recovery}
\titlerunning{Policy-Based Endoscopic Camera Pose Recovery}
%

\author{Jan Emily Mangulabnan\inst{1}\thanks{These authors contributed equally to this work.}
\and
Akshat Chauhan\inst{1}\printfnsymbol{1} \and
Laura Fleig\inst{1} \and
Lalithkumar Seenivasan\inst{1}
\and
Roger D. Soberanis-Mukul\inst{1}
\and
S. Swaroop Vedula\inst{1}
\and
Russell H. Taylor\inst{1,2}
\and
Masaru Ishii\inst{2}
\and
Gregory D. Hager\inst{1}
\and
Mathias Unberath\inst{1, 2}
}
\authorrunning{J.E. Mangulabnan et al.}
%
\institute{Johns Hopkins University, Baltimore MD 21218, USA \and
Johns Hopkins Medical Institutions, Baltimore MD 21218, USA\\
\email{\{jmangul1, unberath\}@jh.edu}}



  
\maketitle              
\begin{abstract}
In endoscopic surgery, surgeons continuously locate the endoscopic view relative to the anatomy by interpreting the evolving visual appearance of the intraoperative scene in the context of their prior knowledge. Vision-based navigation systems seek to replicate this capability by recovering camera pose directly from endoscopic video, but most approaches do not embody the same principles of reasoning about new frames that makes surgeons successful. Instead, they remain grounded in feature matching and geometric optimization over keyframes, an approach that has been shown to degrade under the challenging conditions of endoscopic imaging like low texture and rapid illumination changes. Here, we pursue an alternative approach and investigate a policy-based formulation of endoscopic camera pose recovery that seeks to imitate experts in estimating trajectories conditioned on the previous camera state. Our approach directly predicts short-horizon relative motions without maintaining an explicit geometric representation at inference time.
It thus addresses, by design, some of the notorious challenges of geometry-based approaches, such as brittle correspondence matching, instability in texture-sparse regions, and limited pose coverage due to reconstruction failure.
We evaluate the proposed formulation on cadaveric sinus endoscopy. Under oracle state conditioning, we compare short-horizon motion prediction quality to geometric baselines achieving lowest mean translation error and competitive rotational accuracy. We analyze robustness by grouping prediction windows according to texture richness and illumination change indicating reduced sensitivity to low-texture conditions.
These findings suggest that a learned motion policy offers a viable alternative formulation for endoscopic camera pose recovery.


\end{abstract}
\section{Introduction}

In endoscopic surgery, video inherently encodes motion.
The evolving appearance of the anatomical scene reflects the camera movement that produced it.  
Surgeons actively interpret these visual changes, combining them with their knowledge of anatomy to maintain spatial awareness within narrow corridors. However, intraoperative navigation remains challenging, particularly in complex anatomy with high variability between patients. To assist with navigation, clinicians use tracking systems with preoperative imaging as a reference. While these systems offer an additional spatial context beyond the endoscopic view, they often depend on external tracking hardware which can complicate the clinical workflow.

Vision-based navigation systems have gained interest as an alternative to reduce reliance on external hardware, aiming to recover camera motion directly from endoscopic video. Most existing approaches estimate pose through geometric reasoning, either through explicit feature correspondences in structure-from-motion or SLAM pipelines~\cite{soberanis2025navigation,teufel2024oneslam,liu2022sage,mur2015orb,elvira2024cudasift}, or through learning-based pipelines that predict correspondence or depth~\cite{ozyoruk2021endoslam,guo2025endo3r}. However, these methods degrade under the challenging conditions of endoscopic imaging, where specular reflections, low texture, and repetitive structures destabilize visual signals used for pose estimation~\cite{dunn2024nfl,liu2022sage}. These algorithms operate on the assumption that images are independent observations, decoupling feature extraction and matching from the optimization process.

\begin{figure}[h]
    \centering
    \includegraphics[width=0.6\linewidth]{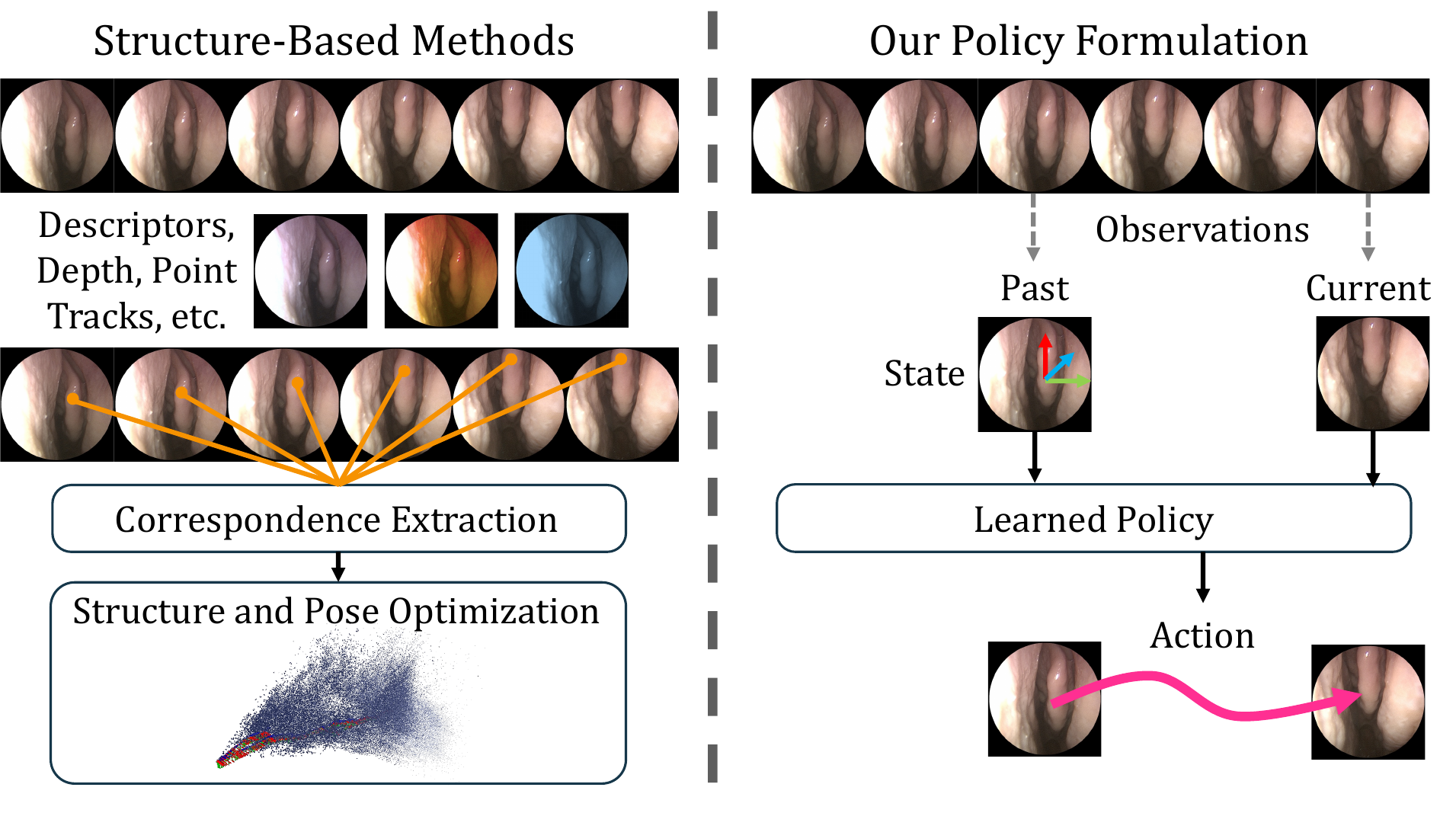}
    \caption{ Conceptual comparison between structure-based pose estimation pipelines and the proposed policy-based formulation.}
    \label{fig:comparison}
\end{figure}

Rather than relying on intermediate geometric understanding, we reconsider camera pose estimation from a motion-centric perspective injecting the assumption that consecutive observations within a video depict the same scene.
Differences in appearance are thus explained by a latent relative pose across a short temporal horizon.
We formulate pose estimation as sequential motion inference conditioned on the previous camera state, where the camera acts as an agent, and predict relative motions to the current observation (Fig.~\ref{fig:comparison}). This allows us to train a policy to imitate incremental expert trajectories, enabling direct motion prediction without an explicit scene representation at inference time.

\subsection{Related Work}
Classical structure-from-motion and SLAM-based systems recover pose by optimizing geometric consistency across frames~\cite{teufel2024oneslam,mur2015orb,elvira2024cudasift}. These methods benefit from enforcing global geometric constraints and maintaining a structural map of the anatomical environment. However, these depend on stable feature correspondences and robust descriptor representations of endoscopic video. Learning-based approaches have been introduced to replace handcrafted descriptors with representations tailored to endoscopic imagery~\cite{liu2020descriptor,vagdargi2025r2d2e,barbed2022superpoint} or incorporate dense correspondence models within reconstruction frameworks~\cite{teufel2024oneslam,lee2025pituitary}. Other approaches leverage scene structure estimated from monocular depth estimation~\cite{liu2022sage,liu2019dense,ozyoruk2021endoslam} or neural rendering pipelines, including neural radiance fields~\cite{shan2023endoslam,shan2024enerf,ruthberg2025neural} and Gaussian splatting-based models~\cite{wang2024endogslam,liu2024endogaussian}. These estimate camera pose through geometric consistency of the reconstructed volume. While these methods improve robustness compared to traditional approaches, they continue to rely on stable depth prediction, correspondence quality, or map consistency.

Policy-based and imitation learning paradigms have gained increasing traction in surgical robotics to learn control policies from expert demonstrations. Recent work predicts instrument or robot motion, typically treating the camera as an external observer of task execution~\cite{furnari2025sequence,he2025surgworld,kim2024surgical}. Large-scale hierarchical systems such as SRT-H similarly map multi-view RGB endoscopic observations to long-horizon surgical policies~\cite{kim2025srt}. Similarly, imitation learning has been explored in X-ray-guided robotic systems to learn task-conditioned control policies, where visual input guides robot actions toward a surgical objective~\cite{klitzner2025xraypolicy}. In these cases, the learned policy predicts task-level control signals. In contrast, we apply imitation learning to endoscopic camera pose estimation. We treat the camera's ego-motion as the prediction target and learn a policy that infers relative camera displacement from visual input.

Based on this perspective, we investigate a policy-based formulation of camera pose estimation, providing a complementary alternative to structure-driven pipelines under challenging endoscopic imaging conditions. In this work we:
\begin{itemize}
    \item \textbf{introduce a policy-based formulation of endoscopic camera pose estimation} that models camera motion as a sequential prediction problem learned from expert trajectories;
    \item \textbf{develop a visual policy conditioned on the current state} to predict the relative camera displacement without maintaining an explicit scene cues during inference; and
    \item \textbf{evaluate performance against structure-driven navigation pipelines,} considering challenging endoscopic imaging conditions.
\end{itemize}

\section{Methods}
\begin{figure}[h]
    \centering
    \includegraphics[width=0.7\linewidth]{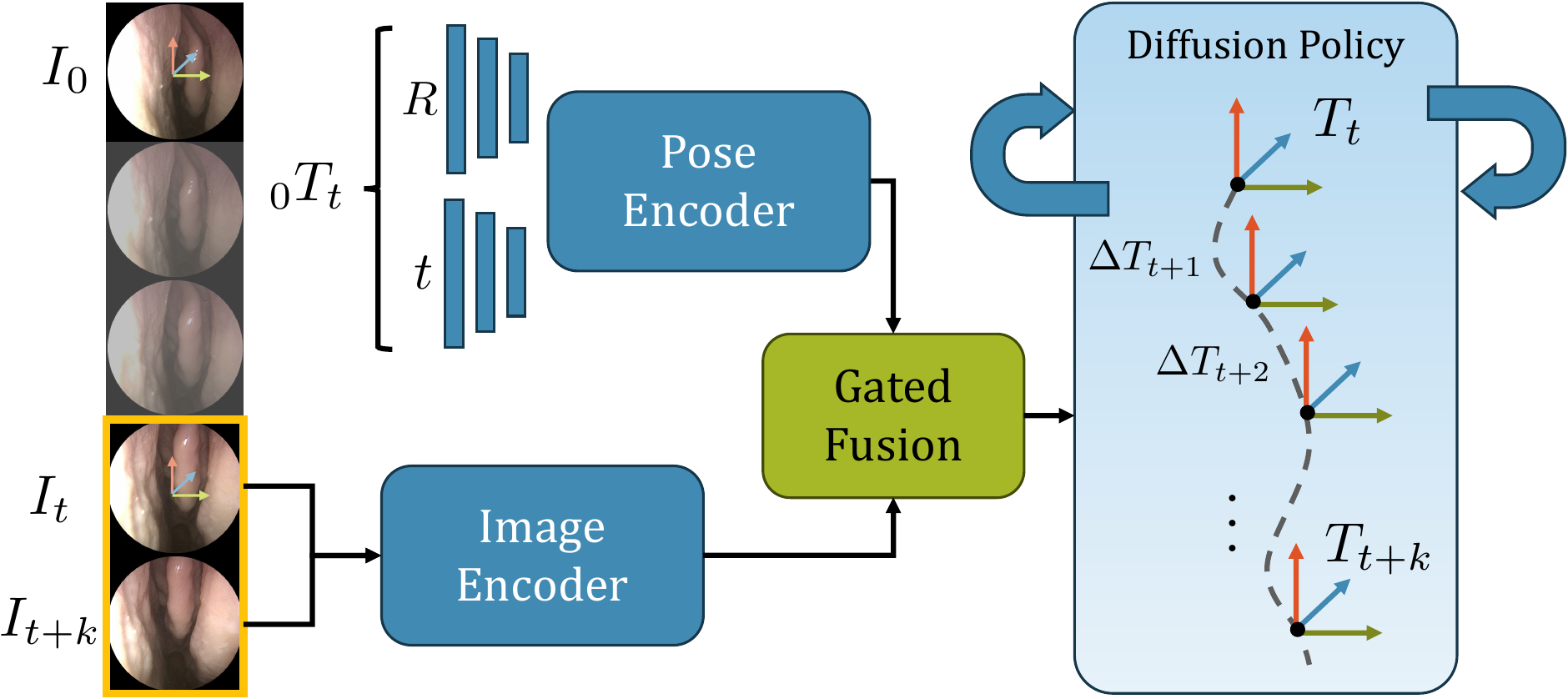}
    \caption{Overview of our proposed policy-based architecture for pose estimation.}
    \label{fig:architecture}
\end{figure}

We formulate endoscopic camera pose recovery as a short-horizon motion inference problem conditioned on a known source state, encoding images and pose to diffuse a sequence of relative motion increments.
An overview of our proposed architecture is shown in Fig.~\ref{fig:architecture}.

\subsection{Problem Formulation}
Given an image, $I_t$ with camera pose $T_t \in SE(3)$, we define an anchored state representation within each video by expressing all poses relative to the first frame of the sequence. This anchoring establishes a consistent coordinate frame to preserve temporal correlation.
For a horizon of $k$ steps, we estimate the relative motion between images $I_t$ and $I_{t+k}$.
Let the incremental motions be denoted as $\Delta \{T_{t+i}\}_{i=1}^k$, where each $\Delta T_{t+i} \in SE(3)$ represents the relative transformation between consecutive frames.
Relative motion is represented as an incremental transformation, $\Delta T_{t+1} = T_t^{-1} T_{t+1}$, and similarly for subsequent steps up to $k$.
This formulates our learned motion policy $\pi_\theta$ that predicts the sequence of relative camera motions conditioned on visual input and the current state:
\begin{equation}
    \pi_{\theta}(I_t, I_{t+k}, T_t) = \{\Delta T_{t+i}\}^k_{i=1}
\end{equation}

We represent camera pose $T_t \in SE(3)$ as a rigid transformation composed of rotation $R_t \in SO(3)$ and translation $t_t \in \mathbb{R}^3$.  Each transformation is parameterized as a 6D vector consisting of a rotation vector representation (axis-angle) and a 3D translation vector. During training, both state and action representations are normalized using statistics computed over the training set to stabilize optimization. The policy predicts a sequence of normalized action vectors, which are subsequently denormalized and composed to recover camera motion in $SE(3)$.

\subsection{Model Architecture}
We adopt the policy-based motion architecture of~\cite{sun2025poseinsert}, which encodes visual observations with current state. Given a source image $I_t$ and target image $I_{t+k}$, the model learns to predict the sequence of relative motions to transition from the previous state to the current observation.
We extract visual features from $I_t$ and $I_{t+k}$ based on the image encoder of~\cite{wen2024foundationpose} for encoding image pairs with shared weights and the previous camera state $T_t$ is encoded separately~\cite{li2019cdpn}. Visual and state features are then fused with through the residual gating mechanism proposed in~\cite{sun2025poseinsert}. This allows the model to jointly reason about the previous state and the current observation to infer the motion that produced the visual change.

The fused representation is passed to a conditional diffusion-based action decoder that predicts a sequence of $k$ relative motions. We model the trajectory as a sequence of 6D deltas $x_0 = \{\Delta \varepsilon_{t+i}\}_{i=1}^k$, where each $\Delta \varepsilon \in \mathbb{R}^6$ parameterizes relative translation and rotation. The diffusion decoder is implemented as a 1D conditional U-Net that operates over the temporal dimension of the motion sequence, predicting denoised relative pose deltas conditioned on the fused image-state representation.

\section{Experiments}

\subsection{Dataset}
We evaluate our approach on video from sinus endoscopy, providing per-frame ground-truth camera poses.
We use an in-house cadaveric sinus endoscopy dataset collected from 12 specimens, comprising of 55 video sequences spanning preoperative, intraoperative, and postoperative anatomy from both left and right nostrils. We split the dataset with 9 specimens for training, 1 for validation, and 2 for test. We obtained ground-truth camera trajectories using an NDI Polaris Hybrid Position Sensor (Northern Digital Inc., Waterloo, Canada) to track an infrared marker configuration rigidly mounted on the endoscope. A standard checkerboard hand-eye calibration procedure was performed to compute the fixed transform between the marker configuration and the endoscopic camera, enabling recovery of per-frame camera poses in a consistent metric coordinate system.

\subsection{Experimental Setup}
Images were undistorted using parameters obtained via checkerboard calibration and resized to 160x160 for training.
Poses were represented in 6D with translation and axis-angle rotation, then normalized independently for both the state and action representations.
The trajectory horizon was set to $k = 8$ action steps.
The policy model was trained using an AdamW optimizer $(\beta_1=0.95, \beta_2=0.999)$ with a learning rate of $1e^{-4}$ and weight decay of $1e^{-6}$. Training was conducted for 
20 epochs, with the best-performing model selected based on the validation set. All experiments were performed on a single NVIDIA GeForce RTX 3060 GPU with a batch size of $32$. For diffusion training, we use a DDIM scheduler with 100 diffusion timesteps and cosine noise schedule, using an $x_0$-prediction objective. During training, we additionally apply zero-mean Gaussian SE(3) noise to the input state pose to improve robustness to compounding rollout errors.

We compare our approach against representative geometric and learning-based pose estimation pipelines. As a classical geometric comparison, we use Structure-from-Motion (SfM) implemented in GLOMAP~\cite{pan2024glomap} with SIFT, then with learned feature descriptors~\cite{liu2020descriptor}. We also compare against a SLAM system that leverages learning-based point tracks for pose estimation~\cite{teufel2024oneslam}. For these baselines, predicted trajectories are aligned to ground truth using a Sim(3) transformation per sequence prior, ensuring fair comparison under scale ambiguity. Our method predicts metric relative motions conditioned on ground-truth state and therefore does not require post-hoc scale alignment.

\subsection{Trajectory Prediction}
We evaluate motion prediction quality under an oracle state assumption to isolate the learned motion from compounding state-estimation errors. In this setting, the ground-truth camera pose $T_t^{gt}$ at time $t$ is provided as the conditioning state. This matches the assumed inference regime of our model, which requires pose input of the previous state. We assess local motion modeling accuracy by reporting Relative Pose Error (RPE). We compare the predicted relative motion over $w$ steps to the ground-truth relative transform:
\begin{equation}
    \Delta T_{t:t+w}^{gt} = (T_t^{gt})^{-1} T_{t+w}^{gt}
    \qquad
    \Delta T_{t:t+w}^{pred} = \prod_{i=1}^{w} \left( \Delta T_{t+i}^{pred} \right)
\end{equation}
We then compute the translation error as the Euclidean distance between predicted and ground-truth relative translations (mm), and rotation error as the geodesic angle between relative rotations (degrees).


\subsection{Robustness Analysis}
As endoscopic imaging presents well-documented challenges for vision-based motion estimation, including low-texture mucosal surfaces and rapid illumination changes, we evaluate robustness with the sinus dataset under these conditions. We compute image statistics to approximate these artifacts for each window $(I_t, I_{t+w})$.
We quantify low-texture regions using the mean Sobel gradient magnitude $s_{texture}$ in the source image. Rapid illumination changes are also measured by computing the change in image intensity, $s_{\Delta illum}$ from image $t$ to $t + w$. We compute these scores considering only valid pixels $n$ in the endoscopic field-of-view where $g(\cdot)$ is the grayscale image and $G_x, G_y$ are Sobel derivatives:

\begin{subequations}
    \begin{align}
        s_{texture}(I_t) &= \frac{1}{n} \sum_{i = 1}^{n} \sqrt{G_x(i)^2 + G_y(i)^2}\\
        s_{\Delta illum}(I_t, I_{t+w}) &= \left| \frac{1}{n} \sum_{i = 1}^n g_{t+w}(i) - \frac{1}{n} \sum_{i = 1}^n g_{t}(i) \right|
    \end{align}
\end{subequations}

\begin{figure}[h]
    \centering
    \includegraphics[width=0.8\linewidth]{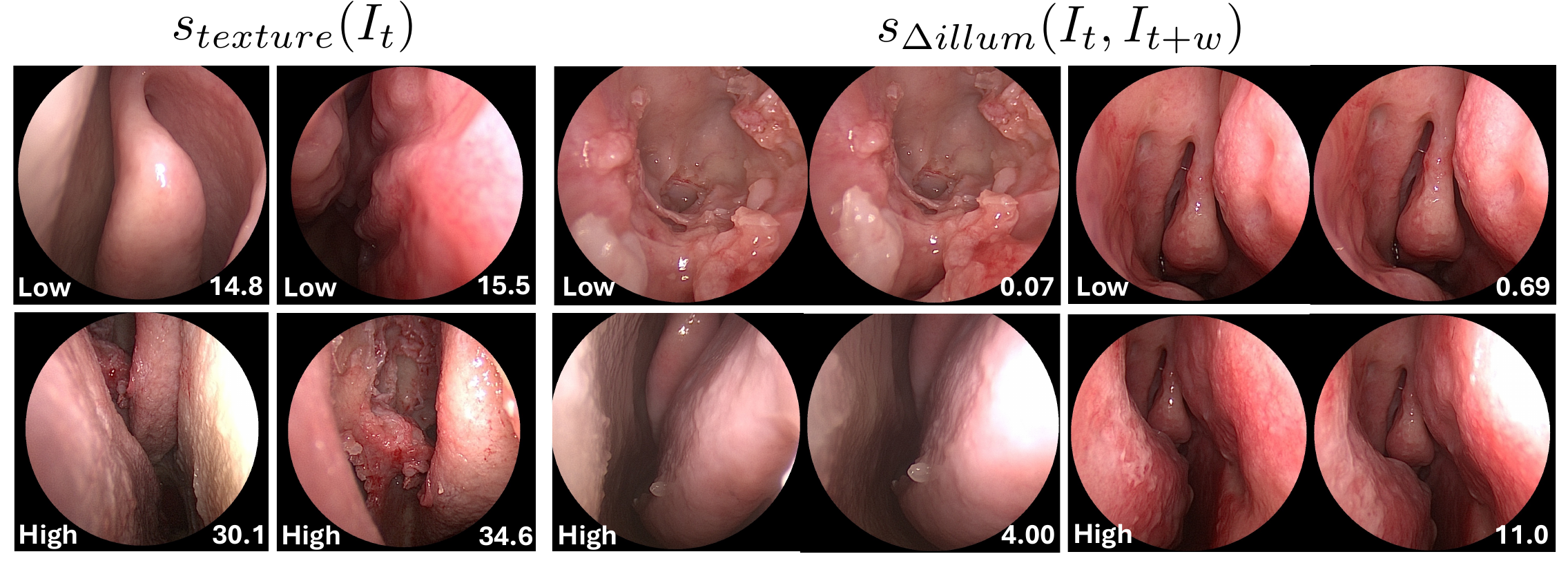}
    \caption{Example images of endoscopic challenge cases for low versus high texture and change in illumination from start to end of the window classified by score.}
    \label{fig:scores}
\end{figure}

The windows $w$ are then grouped into bins based on the global distribution of each difficulty score across the test split, and RPE is reported separately for each bin. We separate these bins into two groups corresponding to lower (25\%) and upper (25\%) quartile ranges for each score. Sample images of these classifications are shown in Fig~\ref{fig:scores}.


\section{Results and Discussion}

\subsection{Trajectory Prediction}
We evaluated short-horizon motion prediction under the oracle state assumption to isolate predicted motions from accumulated state-estimation drift. Performance is measured using window-based relative pose error (RPE) in translation (mm) and rotation (degrees) as reported in Table~\ref{tab:rpe_traj}. We also report pose coverage, defined as the percentage of input frames yielding a valid pose estimate.

\begin{table}[h]
    \centering
    \caption{Window-based Relative Pose Error (RPE) at $w=8$ on the sinus test set. 
    All (6) reports mean $\pm$ std translation (mm), rotation (deg), and pose coverage (\%). 
    Challenge and Other 5 report mean translation / rotation.}
    \label{tab:rpe_traj}
    \begin{tabular}{lccc|cc}
    \toprule
    & \multicolumn{3}{c}{\textbf{All (6)}} 
    & \multicolumn{2}{c}{\textbf{Sequence Breakdown}} \\
    \cmidrule(lr){2-4} \cmidrule(lr){5-6}
    \textbf{Method} 
    & Trans. & Rot. & Cov.
    & Challenge & Other 5 \\
    \midrule
    SfM (SIFT) 
    & $8.60 \pm 16.1$ & $1.66 \pm 2.1$ & $89$ 
    & $44.56 \,/\, 6.32$ & $1.41 \,/\, 0.73$ \\

    SfM (Learned Desc.) 
    & $6.36 \pm 10.2$ & $2.30 \pm 2.4$ & $100$ 
    & $28.98 \,/\, 6.96$ & $1.83 \,/\, 1.37$ \\

    OneSLAM (TAP-based) 
    & $5.40 \pm 5.6$ & $3.06 \pm 1.3$ & $100$ 
    & $17.76 \,/\, 5.64$ & $2.93 \,/\, 2.54$ \\

    \textbf{Ours ($\pi_\theta$)} 
    & $3.70 \pm 3.0$ & $2.21 \pm 1.0$ & $100$ 
    & $10.31 \,/\, 3.99$ & $2.34 \,/\, 1.85$ \\

    \bottomrule
    \end{tabular}
\end{table}

Across all six test sequences, our policy achieves the lowest mean translation RPE and competitive rotational accuracy compared to geometric baselines. Pose coverage remains at 100\% as feature-based SfM exhibits occasional tracking failures. However, the overall mean errors are dominated by a single particularly challenging video sequence that substantially increases the aggregate variance. In this sequence, SfM (SIFT) recovers only 51\% of poses, while the remaining methods maintain near-complete coverage. To better contextualize this effect, we also report a sequence-level breakdown to isolate the effect of this outlier and report performance separately on the remaining sequences. Notably, the proposed policy-based formulation demonstrates lower error than geometric baselines on the challenge sequence, while remaining competitive on the remaining sequences.

\subsection{Robustness Analysis}

While RPE provides a measure of motion prediction quality, endoscopic imaging presents well-known challenges that can disproportionately affect pose recovery. To examine sensitivity to these artifacts, we analyze performance under  image conditions reflecting texture variation and illumination change shown in Table~\ref{tab:robust_all}.

\begin{table}[h]
    \centering
    \setlength{\tabcolsep}{3pt}
    \caption{Robustness analysis at $w=8$ for sinus endoscopy. Windows are globally stratified into low (bottom 25\%) and high (top 25\%) bins for each artifact. Translation RPE is reported in mm (mean $\pm$ std). $\Delta$ denotes the difference between High and Low.}
    \label{tab:robust_all}
    \begin{tabular}{lccc|ccc}
    \toprule
     & \multicolumn{3}{c}{\textbf{Texture}} 
     & \multicolumn{3}{c}{$\Delta$ \textbf{Illum.}} \\
    \cmidrule(lr){2-4} \cmidrule(lr){5-7}
    \textbf{Method} & Low & High & $|\Delta|$ 
                   & Low & High & $|\Delta|$ \\
    \midrule
    SfM (SIFT) 
    & $2.60 \pm 4.2$ & $1.43 \pm 1.1$ & $1.17$
    & $2.16 \pm 6.8$ & $6.07 \pm 26.9$ & $3.91$ \\

    SfM (Learned Desc.) 
    & $2.89 \pm 5.9$ & $1.72 \pm 1.3$ & $1.17$
    & $2.28 \pm 6.5$ & $5.71 \pm 24.5$ & $3.43$ \\

    OneSLAM (TAP-based) 
    & $3.59 \pm 4.5$ & $2.40 \pm 2.3$ & $1.19$
    & $2.96 \pm 6.7$ & $6.10 \pm 24.4$ & $3.14$ \\

    \textbf{Ours ($\pi_\theta$)} 
    & $3.01 \pm 4.0$ & $2.21 \pm 1.4$ & $0.80$
    & $2.71 \pm 6.4$ & $5.86 \pm 24.3$ & $3.15$ \\

    \bottomrule
    \end{tabular}
\end{table}

We compute the performance gap between low- and high-difficulty windows ($|\Delta|$ in Table~\ref{tab:robust_all}) to quantify sensitivity to each artifact. The proposed policy-based formulation exhibits reduced sensitivity to low-texture conditions compared to geometric baselines, reflected by a smaller performance gap. All methods demonstrate similar errors under large illumination changes, suggesting that abrupt photometric variation remains a challenge independent of modeling paradigm. Overall, these results suggest that direct short-horizon motion modeling can alleviate certain feature-drive failure modes, but does not inherently address severe photometric instability.



\section{Conclusion}
We presented a policy-based formulation of endoscopic camera pose recovery that models motion conditioned on the previous image, camera state, and current observation. By reframing pose estimation as the imitation of incremental expert trajectories, our approach predicts short-horizon relative motion without explicit geometric reconstruction at inference time. Under oracle state conditioning, we isolate motion prediction quality from accumulated drift and achieve the lowest mean translation error with competitive rotational accuracy and full pose coverage compared to geometric baselines. Further analysis indicates reduced sensitivity to low-texture conditions, while illumination variability remains a shared challenge across methods. Future work will focus on stabilizing long-horizon rollouts by incorporating intermediate image observations or constraining the sequential composition of relative motion predictions. The sequential $k$-step structure also enables principled uncertainty estimation, as overlapping temporal windows provide multiple independent predictions for the same pose. Thus, their disagreement can serve as a consistency-based confidence measure for drift correction or hybrid geometric refinement.

    



%
%
%
\bibliographystyle{splncs04}
\bibliography{references}
%




\end{document}